\documentstyle[emulateapj,psfig,pstricks,apjfonts]{article}
\newcommand{\etal}{{\it et~al. }}
\newcommand{\eg}{{\it e.g. }}

\newcommand{\Hi}{\ion{H}{1}}
\newcommand{\Hii}{\ion{H}{2}}

\newcommand{\jonezero}{{\sl J} = $1 {\rightarrow} 0$}

\begin{document}

\submitted{For Publication in The Astrophysical Journal: Accepted 29 April 1999}
\title{Nonthermal X-Ray Emission From the Shell-Type SNR G347.3$-$0.5}

\author{Patrick Slane\altaffilmark{1}, 
Bryan M. Gaensler\altaffilmark{2,3,4}, 
T. M. Dame\altaffilmark{1},
John P. Hughes\altaffilmark{5}, 
Paul~P.~Plucinsky\altaffilmark{1}, 
and Anne Green\altaffilmark{3}}

\altaffiltext{1}{Harvard-Smithsonian Center for Astrophysics, 60 Garden Street,
Cambridge, MA 02138}
\altaffiltext{2}{Center for Space Research, Massachusetts Institute of 
Technology, Cambridge, MA 02139}
\altaffiltext{3}{Astrophysics Department, School of Physics A29, University of 
Sydney, NSW 2006, Australia}
\altaffiltext{4}{Australia Telescope National Facility, CSIRO, PO Box 76, 
Epping, NSW 1710, Australia}
\altaffiltext{5}{Department of Physics and Astronomy, Rutgers, The State 
University of New Jersey, 136 Frelinghuysen Road, Piscataway, NJ 08854-8019}

\accepted{April 29, 1999}

\begin{abstract}
Recent ASCA observations of G347.3$-$0.5, an SNR discovered in
the ROSAT All-Sky Survey, reveal nonthermal emission from a region
along the northwestern shell (Koyama \etal 1997). Here we report on
new pointed ASCA observations of G347.3$-$.5 which confirm this result 
for all the bright shell regions and also reveal similar emission, although 
with slightly different spectral properties, from the remainder of the SNR. 
Curiously, no thermal X-ray emission is detected anywhere in the remnant.  
We derive limits on the amount of thermal emitting material present in
G347.3$-$0.5 and present new radio continuum, CO and infrared results which
indicate that the remnant is distant and of moderate age. We show that our
observations are broadly consistent with a scenario that has most of the 
supernova remnant shock wave still within the stellar wind bubble of its
progenitor star, while part of it appears to be interacting with
denser material. A point source at the center of the remnant has spectral
properties similar to those expected for a neutron star and may represent
the compact relic of the supernova progenitor.

\end{abstract}

\keywords{ISM: individual (G347.3$-$0.5) --- supernova remnants --- X-rays:
interstellar}

\section{INTRODUCTION}
While standard wisdom has long held that some significant component of
the cosmic ray spectrum is the result of particle acceleration by shocks
in SNRs, only recently has compelling evidence of such
a process been observed. Studies carried out with the {\it Advanced Satellite
for Cosmology and Astrophysics} (ASCA) 
have clearly established the nonthermal nature of the emission from the
limb-brightened regions of the remnant of SN~1006 (Koyama et al.~1995). 
Both the spectrum
and the morphology of the observed X-rays can be explained by models in which
synchrotron radiation is produced from electrons with energies up to 100~TeV
accelerated by the remnant blast wave (Reynolds 1996), a scenario which is
bolstered by the apparent detection of TeV $\gamma$-rays from SN~1006 (Tanimori
et al.~1998).

X-ray spectral results similar to those for SN~1006 are observed for
the remnant G347.3$-$0.5 (Figure 1),
a bright SNR discovered in the ROSAT All-Sky Survey (RASS; Pfeffermann \&
Aschenbach 1996 -- the ROSAT catalog source name for the remnant is
RX~J1713.7$-$3946). Koyama et al.~(1997) serendipitously observed the northwest
shell of the remnant in a portion of a Galactic Plane survey being conducted
with ASCA and found that the emission from this region appears nonthermal;
the spectrum is featureless, extends to high energies, and is well described by
a power law.

Here we present our new X-ray and radio observations of G347.3$-$0.5
that cover the entire SNR. 
In \S2 below, we describe the several data sets we have for the remnant.
In \S3 we discuss the remnant's distance implied by our CO data and the 
spectral results from the X-ray observations. Section 4 contains a discussion 
of the overall picture the data suggest for G347.3$-$0.5, and the conclusions 
are summarized in \S5.

\vspace{0.2in}
\section{OBSERVATIONS}
Nonthermal emission in supernova remnants is usually associated
with the presence of a central pulsar-powered synchrotron nebula. SN~1006
appears to present a glaring exception to this scenario,
but Cas A also possesses a hard power-law tail
(Allen et al.~1997),
and IC443 harbors a concentration of hard emission (Keohane et al.~1997)
that appears to be associated with a molecular cloud interaction site.
With synchrotron spectra extending to 10~keV and beyond, this implies
the presence of electrons with energies as high as $100-200$~TeV for
typical interstellar magnetic fields.
The origin of this emission is of
considerable significance both in terms of the long-surmised source of
cosmic rays and because it raises questions about what special conditions
are responsible for such production in such a limited number of SNRs.
Like SN~1006 (but unlike Cas A or IC443), the nonthermal
emission from G347.3$-$0.5 {\it dominates} the X-ray flux.
This SNR thus presents significant
evidence for particle acceleration which may lead to a confirmation of
the long-suspected connection with the origin of cosmic rays.
However, in contrast to SN~1006, whose age, distance, and progenitor type are
known, G347.3$-$0.5 is an unfamiliar beast. To understand how and why (or even
if) it is acting as a cosmic ray accelerator, we need to establish the
basic fundamentals of its character.

%%%%%%%% Figure 1 and Figure 2 %%%%%%%%%%%%%%%%
\begin{figure*}[tb]
\pspicture(0,10.5)(18.5,21)

\rput[tl]{0}(0.4,20.7){\epsfxsize=8.5cm
\epsffile{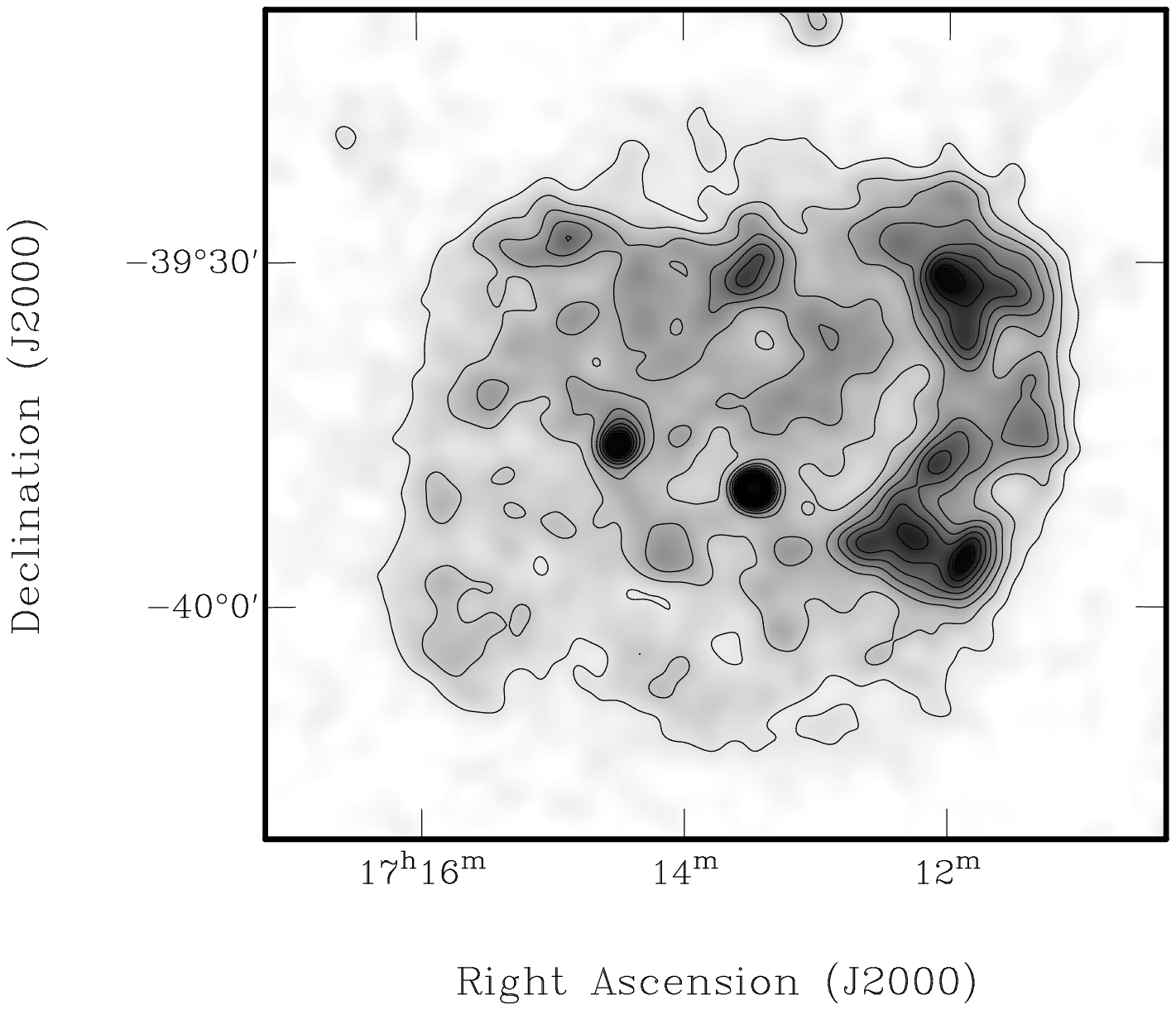}}

\rput[tl]{0}(9.75,20.7){\epsfxsize=8.0cm
\epsffile{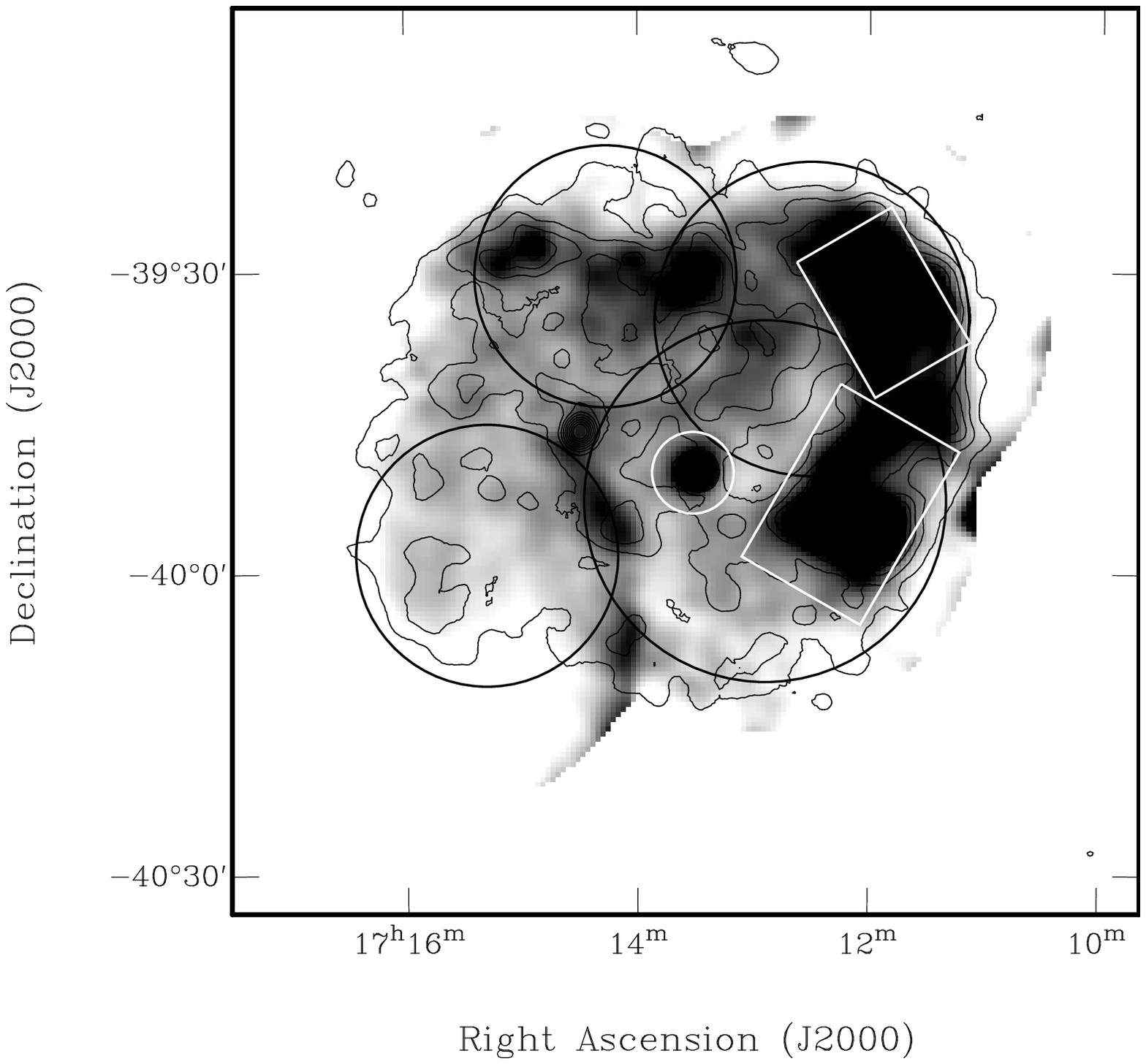}}

\rput[tl]{0}(0,12.7){
\begin{minipage}{8.75cm}
\small\parindent=3.5mm
{\sc Fig.}~1.---
ROSAT PSPC image of G347.3$-$0.5. Two point-like sources are seen
in the central regions. The eastern-most source is associated with a star
while the central source is unidentified. ASCA observations reveal strong
nonthermal emission from the two bright regions along the western limb as
well as weak nonthermal emission from the rest of the interior. Contour levels
start at $1.44 {\rm\ counts\ arcmin}^{-2} {\rm\ s}^{-1}$ and increase inward
in steps of $1.15 {\rm\ counts\ arcmin}^{-2} {\rm\ s}^{-1}$.
\end{minipage}
}

\rput[tl]{0}(9.7,12.7){
\begin{minipage}{8.75cm}
\small\parindent=3.5mm
{\sc Fig.}~2.---
ASCA GIS image of G347.3$-$0.5 with PSPC contours.
Regions used for spectral
studies are indicated. Boxes along the western limb were used for the NW and
SW regions (see text). Two circles from the eastern half of the remnant were
used for the EAST region and a circle around the central point source was used
for the PTSRC region. For deriving upper limits to the thermal emission for
the entire SNR, the four black circles were used with the white circle
excluded.
\end{minipage}
}

\endpspicture
\end{figure*}
%%%%%%%%%%%%%%%%%%%%%%%%%%%%%%%%%%%%%%%%%%%%%%%%%%%%%%%%%%

\subsection{X-Ray}
G347.3$-$0.5 was first discovered in the RASS data (Pfeffermann \& Aschenbach
1996). A follow-up 3~ks pointed observation was then carried out with
the Position Sensitive Proportional Counter (PSPC),
providing a good map of the $\sim 60$~arcmin diameter remnant (Figure 1).
The spectral results were inconclusive, however, being unable to
distinguish between a thermal or nonthermal origin for the soft X-ray
emission.  The 
PSPC also reveals 
two point-like sources in the field of the remnant. The first,
1WGAJ1714.4$-$3945, has a very soft spectrum and appears to be associated with
a star (Pfeffermann \& Aschenbach 1996). We will not
discuss it further. The second source, 1WGAJ1713.4$-$3949,
which is located near the center of the remnant
at position 17$^{\rm h}$~13$^{\rm m}$~28$^{\rm s}$,
$-$39$^{\circ}$~49$^{\prime}$~46$^{\prime\prime}$ (epoch J2000 here
and henceforth), has no reported counterpart. 
The brightest cataloged object near the X-ray source is $\sim 10$~arcsec
away and has an optical magnitude $m_B = 21.5$ ($m_R = 18.4$). 
This separation is consistent with the
expected position errors for the PSPC source given the point spread function of
the detector as well as aspect reconstruction errors. The large X-ray to 
optical flux ratio is consistent with what might be expected for a neutron 
star.  A stellar counterpart is virtually ruled out for such a ratio 
(Stocke et al.~1991), but a background extragalactic source remains a 
possibility.

We observed G347.3$-$0.5 with ASCA on 1997 March 25 for a total of 50~ks
using three pointings of 10~ks duration and one (in the fainter southeast
region) of 20~ks duration, which resulted in complete coverage 
by the Gas-Imaging Spectrometer (GIS). Complimentary chips on
the CCDs were used to image the entire Solid-state Imaging Spectrometer
(SIS) field of view; nevertheless, complete coverage of the remnant was
not possible with the SIS. Risetime  
information for the GIS was sacrificed in order to provide improved temporal
resolution so that timing analysis could be carried out on the central source.
After standard screening procedures for the
GIS and SIS data, we performed spectral fits for several distinct spatial
regions defined as follows: 
NW - a box surrounding the bright region in the northwest which 
was first studied by Koyama et al.~(1997); SW - a box surrounding the bright
region in the southwest; PTSRC - a 3 arcmin radius circle centered on the
central source; and EAST - combined data from two circles in the northeast and
southeast regions of the remnant. 
These regions are shown schematically in Figure 2 where we present the GIS
image. (Also shown are additional circles used in combination with EAST to 
establish limits on the total thermal emission; see \S3.2.2.)
The corresponding spectra are shown in Figure 3. 

%%%%%%%%%%%%%%%% Figure 3 %%%%%%%%%%%%%%%%%%%%%%%%%%%%%%%%%%%%%%%%%
\pspicture(9.5,-1)(18.5,10.3)

\rput[tl]{0}(9.0,10.3){\epsfxsize=8.8cm
\epsffile{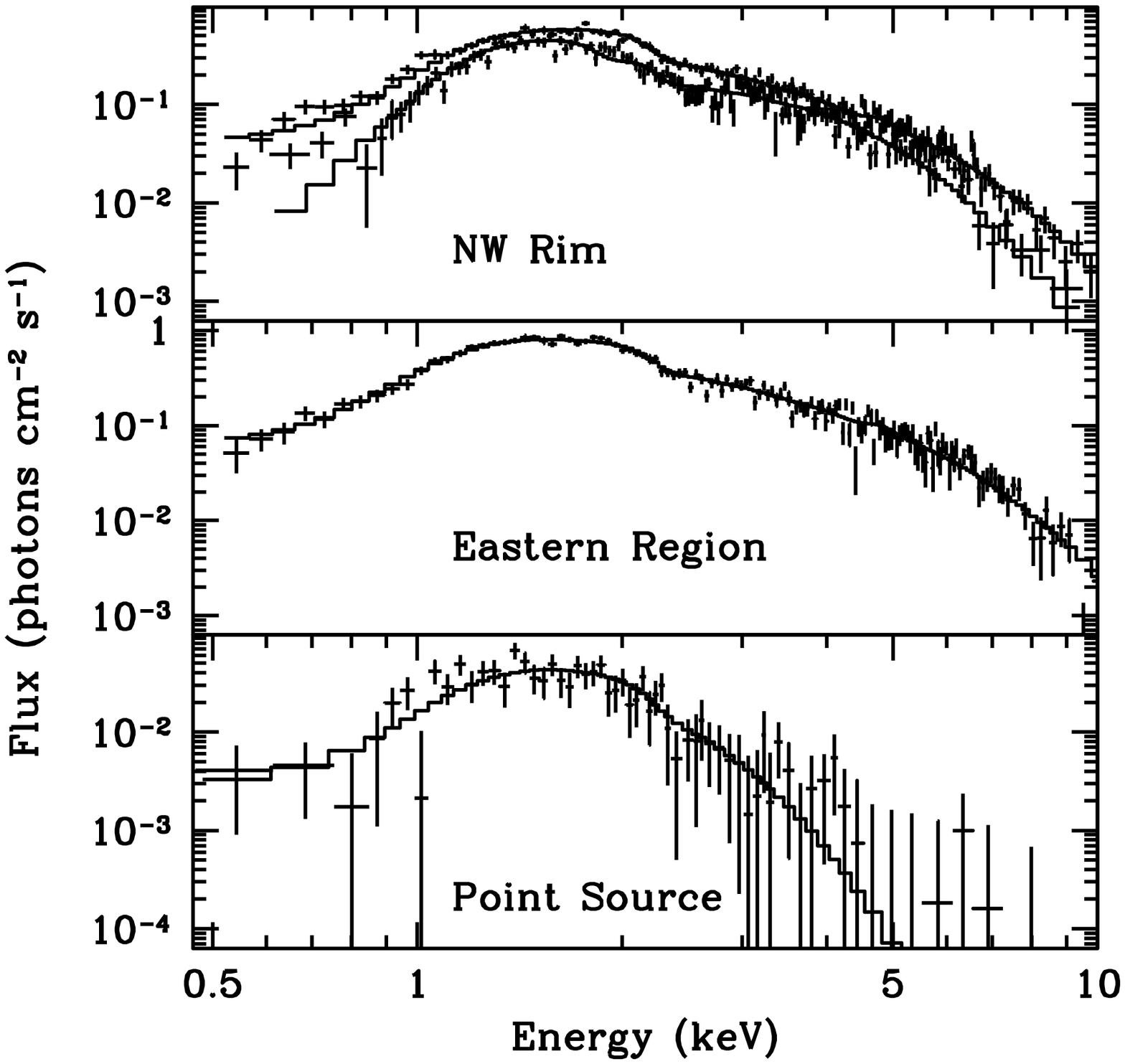}}

\rput[tl]{0}(9.0,1.5){
\begin{minipage}{8.75cm}
\small\parindent=3.5mm
{\sc Fig.}~3.---
ASCA spectra from distinct regions of G347.3$-$0.5 (see
Figure 2) along with best-fit models. Top Panel: GIS (upper) and SIS (lower)
from NW shell. Spectrum is well described by a power law (shown as histogram)
and is indistinguishable from spectrum of SW shell (not shown). Middle Panel:
GIS spectrum from eastern region of remnant along with best-fit power law
model. Lower Panel: GIS spectrum of central point-like source along with
best-fit blackbody model.

\end{minipage}
}
\endpspicture
%%%%%%%%%%%%%%%%%%%%%%%%%%%%%%%%%%%%%%%%%%%%%%%%%%%%%%%%%%%%%%%%%%%

%%%%%%%% Table 1 %%%%%%%%%%%%%%%%%%%%%%
\begin{table*}[t]
\begin{center}
TABLE 1

{\sc G347.3$-$0.5 Spectral Parameters}

\vspace{1mm}
\begin{tabular}{l|lll}\hline\hline
Region & $N_H (\times 10^{21}{\rm\ cm}^{-2})$ & $\Gamma$ (photon) &
$F_x$(0.5-10 keV)\\ \hline
NW Rim & $8.1 \pm 0.4$ & $2.41^{+0.05}_{-0.04}$ & $1.6 \times 10^{-10} {\rm\
erg\ cm}^{-2}{\rm\ s}^{-1}$ \\
SW Rim & $7.9 \pm 0.5$ & $2.40 \pm 0.05$ & $2 \times 10^{-10} {\rm\
erg\ cm}^{-2}{\rm\ s}^{-1}$  \\
Eastern Region & $5.6 \pm 0.4$ & $2.17 \pm 0.05$ & $3 \times 10^{-10} {\rm\
erg\ cm}^{-2}{\rm\ s}^{-1}$ \\ \hline
\end{tabular}
\end{center}
\vspace{-0.5cm}
\end{table*}
%%%%%%%%%%%%%%%%%%%%%%%%%%%%%%%%%%%%%%%

Unlike the line-dominated X-ray spectra typical of SNRs,
the spectra for G347.3$-$0.5 are clearly featureless and extend to at least
10~keV, where the ASCA effective area becomes very small.
For each region, we performed joint fits to data from each GIS detector
as well as to the associated PSPC data from the same spatial region of
the remnant. A power law spectrum provides an
excellent fit to each spectral region of the remnant. The NW and SW regions
display virtually identical spectra while that for the eastern region appears
to have a flatter spectral index and somewhat lower absorption. The best-fit
spectral values are listed in Table 1. Addition of a thermal component (Raymond
\& Smith 1977) to
the spectral model does not result in a significant improvement in the fit;
there is no significant evidence for thermal emission from any portion of the
SNR. We note that a pure thermal bremsstrahlung model (e.g., an
exponential with gaunt factor) yields a considerably 
larger value of $\chi^2$ than does the power law model (e.g., for the
SW region, 
$\chi^2 = 364$ for 293 degrees of freedom compared with $\chi^2 = 312$ for the
power law model).

The SIS data provide higher sensitivity for detection of line features in the
spectra due to its better spectral resolution compared to the GIS. However, due to the extended
nature of the emission, the number of counts in any particular chip is small.
Further, background measurements were possible for only two
of the chips and these were used as representative background for analysis of
data from other chips. Variations in the spectral characteristics of each
chip, plus possible spatial variations in the cosmic and Galactic
X-ray background emission, thus introduce larger uncertainties than
would be obtained with independently measured backgrounds.
We have carried out fits to the SIS data from the NW region,
the SW region, and for the northeast region. In each case, the SIS fits show 
no evidence of spectral lines and the power-law fit parameters are in 
reasonable agreement with those obtained from the joint GIS/PSPC fits. 

The nonthermal nature of the emission from all regions of the remnant is
quite peculiar. Unlike SN 1006, where thermal emission is found in the central
regions, any such component from G347.3$-$0.5 is apparently overwhelmed by
the nonthermal emission. We discuss this surprising result in \S3.

%%%%%%%% Table 2 %%%%%%%%%%%%%%%%%%%%%%%%%%%%%%%%%%
\pspicture(0,-0.5)(8.5,4)
\rput[tl]{0}(-0.5,3.5){
\begin{minipage}{8.75cm}
\small\parindent=3.5mm
\begin{center}
TABLE 2

{\sc Blackbody Model for 1WGAJ1713.4-3949}

\vspace{1mm}

\begin{tabular}{ll}\hline\hline
Parameter & Value \\ \hline
$N_H$ & $5.2^{+1.8}_{-1.6} \times 10^{21} {\rm\ cm}^{-2}$ \\
$kT$ & $0.38 \pm 0.04$ keV \\
$K^{(\rm a)}$ & $25^{+20}_{-10}$ \\
$F_x (0.5-5.0 {\rm\ keV})$ &  $5.3 \times 10^{-12}{\rm\ erg\ cm}^{-2}{\rm\
s}^{-1}$\\ \hline
\multicolumn{2}{l}{(a) $K = (R_{\rm km}/D_{10})^2$ where $R$ is radius of 
emiting}\\
\multicolumn{2}{l}{region in km and $D_{10}$ is distance in units of
10 kpc}\\
\end{tabular}
\end{center}
\noindent

\end{minipage}
}
\endpspicture
%%%%%%%%%%%%%%%%%%%%%%%%%%%%%%%%%%%%%%%%%%%%%%%%%%%%%%%%%%%%%%%%%%%

The spectrum from the unidentified central source 1WGAJ1713.4$-$3949 is
not well constrained by the data. 
Joint fits to the SIS/GIS/PSPC data yield acceptable fits for power-law,
Raymond-Smith, bremsstrahlung, and blackbody
models. The blackbody model 
yields the lowest overall $\chi^2$ fit (though only marginally lower than the
other models), and we summarize these fit parameters
in Table 2 for use in discussing a neutron star scenario for this source
in \S3.2.3. However, we note that additional observations are required
to obtain a higher fidelity characterization of the spectral properties.

%%%%%%%% Figure 4 and Figure 5 %%%%%%%%%%%%%%%%%%%%%
\begin{figure*}[tb]
\pspicture(0,11)(18.5,21)

%\rput[tl]{0}(0.4,20.7){\epsfxsize=8.5cm
%\epsffile{f4.ps}}

%\rput[tl]{0}(9.75,20.7){\epsfxsize=8.5cm
%\epsffile{f5.ps}}

\rput[tl]{0}(0,12.7){
\begin{minipage}{8.75cm}
\small\parindent=3.5mm
{\sc Fig.}~4.---
IRAS 60~$\mu$m/100~$\mu$m ratio image with PSPC contours
overlaid. A marginal enhancement in the ratio appears over the interior
of the remnant, but significant emission is seen outside
the western limb possibly indicating the presence of adjacent cool material.
\end{minipage}
}

\rput[tl]{0}(9.7,12.7){
\begin{minipage}{8.75cm}
\small\parindent=3.5mm
{\sc Fig.}~5.---
MOST image of radio emission at 843 MHz, with PSPC contours.
Emission from the bulk of the SNR rim can be seen with particular enhancements
along the west/northwest regions where bright nonthermal X-ray emission is
seen. Bright emission outside of the western limb corresponds well with the
observed infrared emission from this region.
\end{minipage}
}

\endpspicture
\end{figure*}
%%%%%%%%%%%%%%%%%%%%%%%%%%%%%%%%%%%%%%%%%%%%%%%%%%%%%%%%%%

%%%%%%%% Figure 6 %%%%%%%%%%%%%%%%%%%%%
\setcounter{figure}{5}
\begin{figure*}[b]
%\centerline{\psfig{file=f6.eps,width=18.5cm}}
\vspace{2cm}
\caption{
Total molecular column density over a wide section of the fourth
Galactic quadrant around G347.3-0.5.  Molecular column density, $N(H2)$, was
derived from velocity-integrated CO intensity, $W_{\rm CO}$, using a conversion
factor of $N(H_2)/W_{\rm CO} = 1.9 \times 10^{20} {\rm\ cm}^{-2} {\rm\ K}^{-1}
{\rm\ km}^{-1} {\rm\ s}$. (Strong \& Mattox, 1996).
The CO data are from the surveys of Bronfman et al. (1989) and Bitran et al.
(1997).  The contours are logarithmically spaced at 34, 60, 107, 190, 224, \&
600 $\times 10^{20}$ molecules~cm$^{-2}$.  The lowest contour was 
chosen well above the
instrumental noise ($\sim 9 \sigma$) to emphasize the relatively low molecular
column density toward G347.3$-$0.5 (plus sign).
}
\end{figure*}
%%%%%%%%%%%%%%%%%%%%%%%%%%%%%%%%%%%%%%%%%%%%%%%%%%%%%%%%%%

\subsection{IR and Radio}
We have also investigated the infrared emission from the region with the
Infrared Astronomical Satellite (IRAS) 
60$\mu$m/100$\mu$m ratio (Figure 4).  An enhancement of this ratio,
indicative of hot dust, has been shown to be a powerful discriminator of
remnant emission against the background emission of the Galactic plane
(Saken, Fesen, \& Shull 1992; Slane, Vancura, \& Hughes 1996).  The infrared
emission in the inner Galactic disk is intense and complicated, but as
Figure 4 shows, there is a clear enhancement of the 60 mm/100 mm ratio
coincident with the remnant.  The strongly enhanced ratio to the northwest
of the remnant is due to the \Hii\ region G347.61+0.20 (Lockman 1979), which
may be associated with the same molecular cloud as the remnant (see \S3.1).

Radio maps from the Parkes-MIT-NRAO (PMN) 
6~cm survey (Condon, Griffith, \& Wright 1993)
reveal complex emission from the region surrounding the remnant,
but we find very good morphological correspondence with the infrared emission,
particularly for the regions directly west and northwest of the remnant.
To better delineate any radio emission corresponding to the X-ray remnant,
we have carried out 843 MHz observations
using the Molonglo Observatory Synthesis Telescope (MOST; Robertson 1994).
The synthesis was carried out on 1997 May 27, with the telescope in its Wide
Field mode of operation (Large et al. 1994).
The resulting map (Figure 5)
reveals regions of faint emission which extend along most of
the SNR perimeter, with the most distinct emission corresponding to the bright 
X-ray regions in the northwest and southwest.
Considerable contributions from sidelobes, baseline errors, and grating rings
from surrounding sources result in a rather high noise level of
$\sim 4$~mJy~beam$^{-1}$; the contrast has been adjusted considerably to show
the faint shell emission.

The brightest regions of emission within the confines of the X-ray contours
are found along two arcs in the northwest. These are concentrated along the
edges of the brightest X-ray emission. Arc 1 is centered at approximately 
17$^{\rm h}$~12$^{\rm m}$~46$^{\rm s}$,
$-$39$^{\circ}$~25$^{\prime}$~49$^{\prime\prime}$.
We estimate a surface brightness
range of $\sim 17 - 27 {\rm\ mJy\ beam}^{-1}$ ($27 - 43 {\rm\ mJy\ 
arcmin}^{-2}$) with $\sim 10\%$ uncertainty in these values. Arc 2, centered 
at 
17$^{\rm h}$~11$^{\rm m}$~27$^{\rm s}$,
$-$39$^{\circ}$~32$^{\prime}$~36$^{\prime\prime}$ 
has a 
surface brightness range $\sim 56 - 82 {\rm\ mJy\ beam}^{-1}$ ($90 - 130 
{\rm\ mJy\ arcmin}^{-2}$) with similar uncertainty. This arc appears to be 
most closely correlated with the bright X-ray emission and, although
its east-west elongation is curiously askew with respect to the
perceived SNR shell at this location, the radio filament does overlay
a filament of X-ray emission at this position.
An upper limit to any radio counterpart to the unidentified X-ray point source
1WGA~J1713.4$-$3949 is 15 mJy ($5 \sigma$).
The intense emission beyond the NW limb of the remnant corresponds to the same
\Hii\ region, G347.61+0.20, that was evident in the IRAS map (Figure 4).

Additional radio observations of G347.3$-$05 with higher resolution
and sensitivity have been 
carried out 
with the Australia Telescope Compact Array. Analysis of these data is in 
progress and will be reported in a future publication.

\subsection{CO Line Measurements}
 To further constrain our picture of the environment surrounding the
remnant, we have used observations of the \jonezero\ rotational
transition of CO (at 115 GHz), the best large-scale tracer
of interstellar molecular gas, from the CfA-Chile 1.2 m telescope (Bronfman
et al.\ 1989). Figure 6 is a map of total molecular column
density over a wide section of the inner Galaxy centered on the SNR.  It is
noteworthy that the SNR coincides very closely with the largest and deepest
hole in the molecular column density near the Galactic plane in the inner
fourth quadrant.
Since the CO emission in the direction of the remnant is complex and
extends over $\sim 200{\rm\ km\ s}^{-1}$ in radial velocity, it is
difficult to  determine
which individual molecular cloud or clouds, if any, along the line of
sight might be associated. 
The most likely candidates based on the morphology of the CO and X-ray
emission are a pair of clouds at $-94 {\rm\ km\ s}^{-1}$ and a third at 
$-69 {\rm\ km\ s}^{-1}$ (Figure 7).  
Given that giant molecular complexes with active star formation
sometimes have total velocity extents of ~30 km s$^{-1}$ (e.g., W44;
Dame et al.\ 1986), it is possible that all three of these clouds are
part of the same 
molecular complex that gave birth to the \Hii\ region G347.61+0.20 and the
SNR, with one or more of the clouds significantly perturbed in velocity. 
The cloud at $-69 {\rm\ km\ s}^{-1}$ abuts the region of strongest 
X-ray and radio enhancement, suggesting that the remnant may be expanding 
into this cloud. It is worth noting that a compact region of 
anomalous-velocity emission extending
to $-200 {\rm\ km\ s}^{-1}$ is seen in both 21~cm (Burton 1985) and CO
(Bronfman et al.\ 1989) emission along the line of sight to the remnant.

%%%%%%%%%%%%%%%% Figure 7 %%%%%%%%%%%%%%%%%%%%%%%%%%%%%%%%%%%%%%%%%
\pspicture(-0.5,11.7)(8.5,24)

\rput[tl]{0}(-0.75,24){\epsfxsize=8.8cm
\epsffile{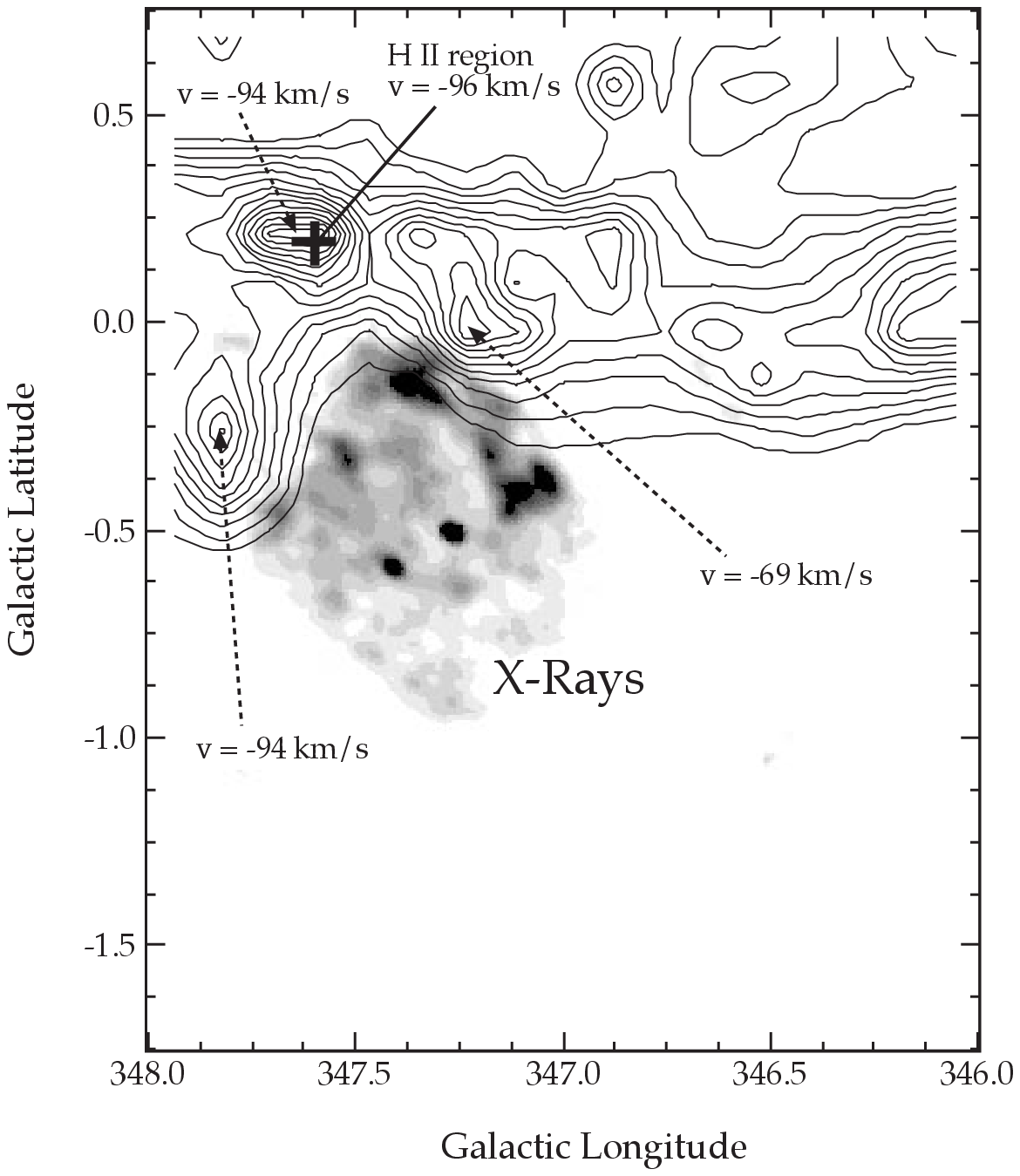}}

\rput[tl]{0}(-0.75,14.5){
\begin{minipage}{8.5cm}
\small\parindent=3.5mm
{\sc Fig.}~7.---
Contours of CO intensity integrated over the velocity range $-100$
to $-50 {\rm\ km\ s}^{-1}$, from the survey of Bronfman et al. (1989).
The contours are spaced every $5 {\rm\ K\ km\ s}^{-1}$, starting at
$15 {\rm\ K\ km\ s}^{-1}$.  The ROSAT PSPC image is
shown in grayscale. The molecular clouds that seem most likely to be
associated with G347.3$-$0.5 are labeled with their LSR velocities.
The \Hii region G347.61+0.20 is marked by a ``+'' sign.
\end{minipage}
}
\endpspicture
%%%%%%%%%%%%%%%%%%%%%%%%%%%%%%%%%%%%%%%%%%%%%%%%%%%%%%%%%%%%%%%%%%%

Given their proximity both in projected position on the sky and in
velocity, the pair of clouds at  
$-94 {\rm\ km\ s}^{-1}$ are almost certainly associated with each other, 
and with the \Hii\ region G347.61+0.20, which has a recombination-line 
velocity of $-96 {\rm\ km\ s}^{-1}$ (Lockman 1979).  An association between 
the SNR and the pair of clouds at $-94 {\rm\ km\ s}^{-1}$ is suggested by 
the presence of massive star formation in one of them, as evidenced by the 
\Hii\ region, and by an enhancement of the CO(2--1)/CO(1--0) line ratio in 
the other.  
Studies of molecular clouds in the Galaxy find values generally below
1 for the ratio $R=W_{\rm CO(2-1)}/W_{\rm CO(1-0)}$.  Sakamoto et al.\
(1995) find a mean value of $0.66$ with a variation from 0.5 to
0.8 that has some dependence on Galactocentric distance. Chiar et
al.~(1994) find a value of $0.85\pm0.63$ from their study of molecular
clouds in the Scutum arm of the Galaxy.  On the other hand 
molecular clouds that appear to be undergoing interactions with SNRs
(e.g., W44 and IC443; Seta et al.\ 1998), show an enhanced
$R$ ratio ($>$1.2) that is indicative of shocked material in the molecular gas.
We obtained CO(2--1) spectra over the range $l=346.5$ to 348.5,
$b=-0.5$ to 0.5, on a 0.25 degree grid (Handa \& Hasegawa 1999,
private communication) from the University of Tokyo
0.6 m telescope on la Silla, Chile, which has the same angular
resolution at the CO(2--1) line as the CfA-Chile 1.2 m at the CO(1--0)
line. Over the range for which CO(2--1) spectra were available, the mean $R$
value was determined to be $0.60 \pm 0.37$.  
However, the largest values of $R$ were associated with the
cloud at $-94 {\rm\ km\ s}^{-1}$: in particular, toward $l=348$,
$b=-0.25$, a value of $R \sim 2$ is found near $-85 {\rm\ km\ s}^{-1}$, in
the wing of a strong (4 K) and well-defined CO(1--0) line near
$-94 {\rm\ km\>s}^{-1}$.  Such an enhancement in the wing of the
CO line is suggestive of an interaction with the SNR shock wave
(Seta et al. 1998).
We note that no enhanced $R$ value is observed for the cloud at $-69 
{\rm\ km\ s}^{-1}$ which appears along the line of sight to the brightest
X-ray limb. However, the CO(2--1) data are coarsely spaced (every 1/4 degree);
more sensitive and closely spaced spectra in this region may reveal such
an enhancement and are of considerable interest.

\section{ANALYSIS}
\subsection{Distance to G347.3$-$0.5}

Koyama et al.\ (1997) compare their measurement of the column density
toward G347.3$-$0.5 (which is in good agreement with the values we
report here) with the total $N_H$ in the direction of the Galactic
center to estimate a distance of 1~kpc to the remnant. We argue that a
distance as small as this is unlikely, based on several lines of
evidence.  Like Koyama et al.\ (1997) we first consider the distance
estimate based on the X-ray derived column density.  The total
line-of-sight column density in the direction of G347.3$-$0.5 is $N_H = 1.2
\times 10^{22}{\rm\ cm}^{-2}$ based upon \Hi\ observations (Dickey \&
Lockman 1990) and confirmed by CO measurements (Bronfman et al.\
1989). This relatively low value is consistent with the presence of a
hole in the molecular column density in this direction mentioned
earlier (see also Fig 6). The $N_H$ value we measure for G347.3$-$0.5
($\sim$$8\times 10^{21}{\rm\ cm}^{-2}$, see Table 1) is a significant
fraction of the total column through the Galaxy and so the distance to
the remnant must be considerably more than 1 kpc.

We have just argued in the preceeding section that the remnant, the
\Hii\ region, and the complex of three molecular clouds in this
direction are all likely to be physically associated. This provides an
opportunity to obtain a quantitative estimate for the distance to the
remnant.  From the rotation curve of the Galaxy (Burton 1988), the
pair of clouds at $-94 {\rm\ km\ s}^{-1}$ lie at a distance of $D =
6.3 \pm 0.4$~kpc or $D = 10.4\pm 0.4$~kpc. Whether the near or far
kinematic distance is appropriate can be resolved in favor of the
former by noting that molecular absorption measurements place the
associated \Hii\ region at a distance of 6.3~kpc (Lockman 1979,
adjusted to the Burton 1988 rotation curve and $R_0 = 8.5$~kpc). 
The cloud at $-69 {\rm\ km\
s}^{-1}$ has a near kinematic distance of 5.4 kpc, assuming that its
radial velocity is purely due to Galactic rotation.  Therefore we
adopt a distance of 6 kpc for the distance to G347.3$-$0.5, with the
warning that this is uncertain by at least $\pm$1 kpc.  In the
following we present results in terms of $D_6 = D/6\, {\rm kpc}$.

\subsection{X-Ray Emission Characteristics}

\subsubsection{Nonthermal Emission}
The nonthermal X-ray emission from G347.3$-$0.5 strongly indicates that the
SNR shock is accelerating particles to very high energies. 
The electron energy associated with synchrotron photons radiated at 
energy $E_x$ is 
$$ E_e \approx \frac{300{\rm\ TeV}}{B_{\mu \rm G}^{1/2}} \left(\frac{E_x}{\rm 
1\ keV}\right)^{1/2}.$$
With observed photon energies up to of 10~keV, the spectrum for G347.3$-$0.5
thus indicates the presence of electrons with energies as high as 300~TeV
for typical magnetic fields of $\sim 10 \mu$G. If protons are accelerated
to similar energies, it would appear that G347.3$-$0.5 may be accelerating cosmic
rays up to energies near the ``knee'' in the cosmic ray spectrum at $\sim
10^{15}$~eV. 
The total energy associated with the X-ray producing electrons is of order
$10^{45} D_6 B_{\mu \rm G}^{-3/2}$~erg which is a small fraction of the
mechanical energy released in the supernova explosion. 

As noted in Table 1, the spectral index for the emission from the bright 
regions along the western limb of G347.3$-$0.5 appears to be steeper than that
for the rest of the diffuse emission from the SNR. The column density from
the spectral fits also differs between the two regions. When $N_H$ is fixed
to the value determined for the NW region, the photon index derived for 
the eastern region is $\Gamma = 2.4$, the same as that for 
the regions along the
western limb. However, the $\chi^2$ for the fit is considerably higher and
the residuals clearly indicate difficulty with such a high absorption value
for the eastern region. 
A variation in column density is not hard to reconcile;
many SNRs exhibit such spatial variations in the foreground absorbing 
column (e.g., Roberts et al.\ 1993). 
For G347.3$-$0.5 this interpretation is particularly reasonable, since the
regions indicating  
higher column density (i.e., the NW and SW enhancement regions) lie closest to 
the Galactic plane (see Figure 7).
Variations in spectral index are also well-understood for synchrotron nebulae
where the finite radiating lifetimes of electrons accelerated by a central
pulsar result in a steepening of the spectral index with distance from the
electron injection point (e.g., Kennel \& Coroniti 1984;
Slane, Bandiera, \& Torii 1998). For synchrotron nebulae the
steepening of the spectrum goes along with a decrease in brightness.
For G347.3$-$0.5, however, no obvious injection point
is evident, and the steeper index appears to be concentrated along the bright
western limb. A synchrotron lifetime argument thus appears problematic in
that the brightest synchrotron region (i.e., the western limb) would
also have to be ``older.''  For completeness we note that the 
nonthermal X-ray emission from SN~1006, with a photon index $\Gamma = 2.95
\pm 0.20$, is steeper than spectra measured here for G347.3$-$0.5.

\subsubsection{Thermal Emission}

Although the bright rim regions of SN~1006 are dominated by synchrotron
emission, ASCA studies of the interior reveal the line-dominated X-ray
thermal component expected
from the hot shocked gas characteristic of SNRs (Koyama et al.\ 1995). 
Our ASCA observations of G347.3$-$0.5 reveal no such component.
The lack of measurable thermal emission
sets limits on the amount of hot material, and thus on the
density of the region in which the SNR evolved. This is illustrated in 
Figure 8 
where we have plotted upper limits to the thermal emission measure
as a function of temperature based upon our spectral analysis. 
For the curve plotted as the solid histogram in Figure 8,
we have combined data from the GIS detectors using regions that cover
most of the remnant (but not the bright regions along the western limb --
see Figure 2) and determined 3-sigma upper limits 
to the amount of thermal emission, assuming normal solar composition,
that could be added to the model. 
The resulting normalization for this thermal component was then scaled by
a factor of two to account for the incomplete coverage of the remnant
that our spectral extraction regions gave. As indicated
in Figure 2, the spectral regions actually cover more than 50\% of the
area of the SNR,
so the upper limits shown are quite conservative.

Under some assumptions about the three-dimensional geometry of the
remnant, the upper limit on the emission measure can be converted into
a limit on the mean density of the ambient medium. For G347.3$-$0.5 we
assume a spherical shell geometry with a measured angular radius of
$\theta_r = 30^\prime$ and an estimated shell thickness equal to
$\theta_r/23$. (Note that the factor 23 was chosen for numerical
agreement with the integrated emission measure for a Sedov solution.)
For the range of allowed emission measures, we then calculate that the
hydrogen number density of the ambient preshock medium in the strong
shock limit (i.e., assuming the 
density jump across the shock front is a factor of 4) is in the
range $n_0$ = 0.014--0.28 $D_6^{-1/2}$ cm$^{-3}$.  The lower bound on
the ambient density corresponds to temperatures of $\sim$0.75 keV,
while the upper bound corresponds to the lowest temperature bin in
Figure 8.

%%%%%%%%%%%%%%%% Figure 8 %%%%%%%%%%%%%%%%%%%%%%%%%%%%%%%%%%%%%%%%%
\pspicture(9.5,8)(18.5,20)

\rput[tl]{0}(9.0,20){\epsfxsize=8.8cm
\epsffile{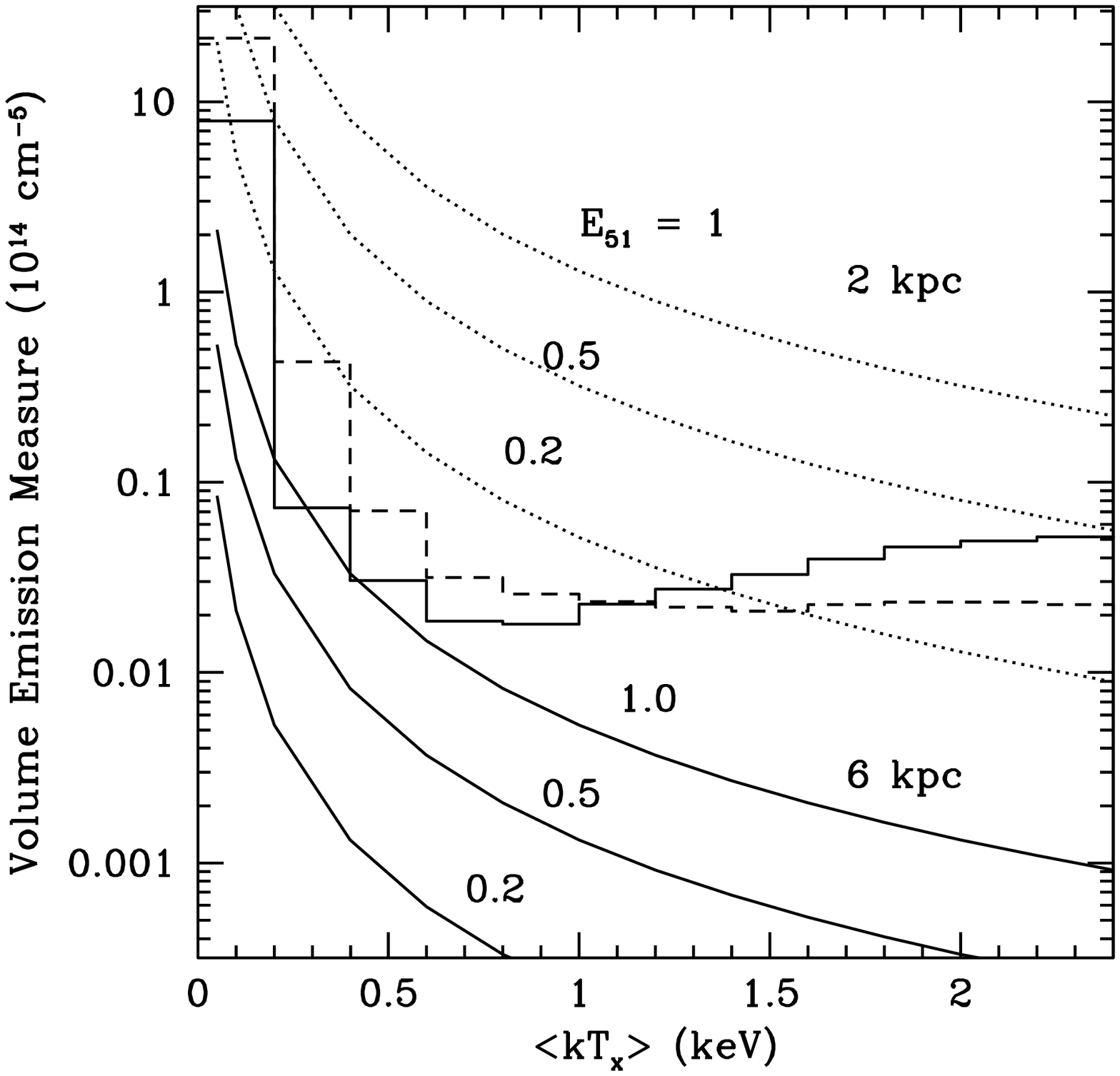}}

\rput[tl]{0}(9.0,11.5){
\begin{minipage}{8.75cm}
\small\parindent=3.5mm
{\sc Fig.}~8.---
Upper limit to derived emission measure
associated with a thermal emission component plotted as a function of
temperature. The two histograms correspond to limits derived for the
entire SNR (solid) based upon scaling results which exclude the bright regions
(see text), and for the bright northwest region (dashed). Also shown are curves
for models derived assuming a Sedov solution. Solid (dotted) curves correspond
to a distance of 6~kpc  (2~kpc) with $E_{51}$ varying from 0.2 (lower) to
1.0 (upper).

\end{minipage}
}
\endpspicture
%%%%%%%%%%%%%%%%%%%%%%%%%%%%%%%%%%%%%%%%%%%%%%%%%%%%%%%%%%%%%%%%%%%

Also plotted in Figure 8
are model curves for explosion energies 
$E_{51} = 0.2$, 0.5, and 1.0 (in units of $10^{51}$ erg)
assuming a distance of 6~kpc (solid curves) and 2~kpc (dotted curves)
under the assumptions that the remnant is in the Sedov (1959)
phase of evolution and the electrons and ions share the same
temperature behind the shock front.
We see that the derived upper limits are not inconsistent with the model 
curves for our prefered distance of 6~kpc. That is, for the range of
parameter space shown, 
the lack of detectable thermal emission is expected for a Sedov-phase SNR;
as with SN~1006, the
thermal emission simply pales in comparison to the bright nonthermal emission.
For such an interpretation, however, the density of the ambient medium
must be very low, as we found above. 

The derived upper limits on emission measure appear to rule out a
Sedov solution for G347.3$-$0.5 if the remnant is at a distance of
2~kpc.  The only acceptable solution requires an unusually low
explosion energy and a rather hot remnant, $kT_x \gtrsim 1.5$~keV.
However this conclusion is sensitive to the assumption of temperature
equilibration between electrons and ions in the post-shock gas.  The
alternative scenario, wherein the electrons and ions acquire the same
mean thermal velocity behind the shock front and then share energy
slowly through Coulomb collisions (see Spitzer 1978), will result in
the mean electron temperature in the remnant ($kT_x$) lagging
considerably behind the post-shock temperature of the ions.  In this
situation the measured electron temperature does not reflect the
actual shock velocity and, in fact, it has been shown (Cox \& Anderson
1982) that the electron temperature has an extremely weak
dependence on the shock temperature.  Furthermore, because  the
equilibration timescale is inversely proportional to the electron
density, the effects of non-equilibration are greatest for low density
remnants, which is precisely the situation we have here for
G347.3$-$0.5.  Although a detailed analysis of the effects of
non-equilibration will be addressed in a future article, we make the
following brief comments here. The curves for the Sedov solution shown
in figure 8 are substantially modified if equilibration on Coulomb
collisional timescales is assumed. First of all, there are no high
temperature solutions for reasonable values of the explosion energy
($E_{51} < 2$) anywhere in the range of emission measures plotted.
Second, it is not possible to rule out a Sedov solution as a model for
the remnant if it is at a distance of 2 kpc.  For example, for this
distance and a mean temperature of $kT_x = 0.5$ keV, emission measures
in the allowed range 0.01--$0.03\times 10^{14} \,\rm cm^{-5}$ predict
ranges for the explosion energy of $E_{51} = 1.4$--2.2 and the
remnant's age of 2100--3400 yrs.  Lower values of the mean temperature
are also acceptable and result in a broad range of possible dynamical
states that, in general, encompass smaller values of the explosion
energy and larger values of the age. For a distance of 6 kpc, a
reasonable dynamical state for G347.3$-$0.5 is possible for a mean
temperature of $kT_x = 0.3$ keV and emission measures in the allowed
range 0.003--$0.07\times 10^{14} \,\rm cm^{-5}$.  Within this range
the explosion energy varies in the range $E_{51} = 2.2$--1.7 and the
remnant's age lies between 18,700 yrs and 40,600 yrs.  So in
conclusion to this discussion, the lack of detected thermal emission
from G347.3$-$0.5 can be reconciled with a broad range of reasonable
dynamical evolutionary states, possible remnant ages, and distances as
long as the mean ambient density around the remnant is low.
This would appear incompatible with the
environment in the vicinity of a molecular cloud unless the remnant has
involved in a cavity produced by the stellar wind of the progenitor star.
We discuss this further in \S4.

The limits on thermal emission used for the preceeding analyses
represent an average of the faint parts of the remnant.  Clearly the
limits will be higher in the bright regions of the SNR.  In the upper
panel of figure 8, the dashed histogram represents the upper limit for
thermal emission from the bright northwestern region of the remnant
using the spectral extraction region illustrated in figure 2. Treating
the emission region as a simple slab through the remnant, the upper
limit at the low end of the temperature scale limits the preshock
density of material at this position to $n_0 < 1 D_6^{-1/2} f^{-1/2}
{\rm\ cm^{-3}}$ where $f$ represents the fraction of the slab volume
that is actually responsible for the emission. This limit suggests
that the ambient density toward the northwest could be considerably greater

\subsubsection{Central Source}
The best spectral fit for the unidentified central X-ray
source 1WGAJ1713.4$-$3949  based upon a joint fit to the GIS,
SIS, and PSPC is obtained for a blackbody spectrum with 
$kT = 0.38 \pm 0.04$~keV (see Table 2). The normalization yields
a relatively small emitting area with radius $R = 3.0^{+1.0}_{-0.7} D_6$~km,
suggesting a compact object. The lack of an optical
counterpart supports the interpretation that the source is a neutron
star associated with G347.3$-$0.5. However, the blackbody temperature is rather
high for a cooling neutron star. Temperatures similar to that observed are
often associated with emission from heated polar caps, but in the case
of 1WGAJ1713.4$-$3949 the emitting area $A$ is considerably
larger than that observed from other sources (\eg PSR 1919+10; Yancopoulos,
Hamilton, \& Helfand 1994),
implying that 1WGAJ1713.4$-$3949 may be a rapidly spinning object 
($A \propto P^{-1}$, where $P$ is the rotation period).

There is no evidence of long-term flux variability between the ROSAT and
ASCA observations (although uncertainties in the spectral model do not
allow us to rule out small variations). We extracted a total of 1100 
counts in the GIS data from the region around the source for use in
timing analysis. 
A search for pulsations for periods longer than $\sim 30$~ms (the highest time
resolution for the entire data set) yielded negative results implying a pulsed 
fraction upper limit of $\sim 55\%$ for the extracted events assuming a 
sinusoidal light curve (Vaughan et al.\ 1994). Based upon fits to the
background-subtracted spectrum for the source, $\sim 47\%$ of the events
in the extraction region were attributable to the source. The derived 
upper limit to the source pulsed fraction is thus of order 100\% and offers
no meaningful constraint.

We note that adequate spectral fits for the central source emission can also
be obtained for power law and Raymond-Smith models. For the power law
model, the spectral index is extremely steep (photon index 
$\Gamma = 4.7^{+0.5}_{-0.4}$) and the column density is large
($N_H = (1.4 \pm 0.2) \times 10^{22}{\rm\ cm}^{-2}$).
For the thermal plasma model, the best-fit column density is $N_H = (1.8 \pm
0.2) \times 10^{22}{\rm\ cm}^{-2}$ which is comparable with the total
absorption through the Galaxy in this direction. 
Thus it is possible
that the source is extragalactic, although the steep X-ray spectral
slope and the lack of radio emission makes this interpretation
somewhat implausible.

\section{DISCUSSION}

Based upon all observations to date, a complete picture of
G347.3$-$0.5 is only just beginning to unfold. Initial X-ray
observations (Pfeffermann \& Aschenbach 1996) 
indicated a nearby ($\sim 1$~kpc) remnant 
which was subsequently suggested as a counterpart to
a guest star from AD393 (Wang, Qu, \& Chen 1997).  The latter authors
argue that the limited information available for such a scenario would
seem to suggest a Type Ib/Ic progenitor explosion. They further
suggest that the morphology of the remnant may show some evidence of a
central ring-like structure which could be associated with evolution
in a wind-blown bubble.  We see no evidence of such structure in
either the original PSPC maps of G347.3$-$0.5 nor in our maps obtained
with ASCA. Further, as argued above, a distance as small as 1~kpc appears
to be very unlikely.

The ASCA spectra for G347.3$-$0.5 clearly indicate the presence of very high
energy electrons, providing evidence for direct acceleration of particles by
the SNR shock. Reynolds (1998) provides a thorough discussion of the maximum
energy constraints for diffusive shock acceleration in SNRs and produces model
results for the nonthermal radiation assuming a constant magnetic field and
explosion characteristics typical of Type Ia progenitors. The maximum energy
for the accelerated particles is determined by: (1) radiative losses by the
particles in the ambient magnetic/photon field; (2) the finite 
acceleration time associated with the SNR age; and (3) particle escape due to
the weakening or absence of magnetohydrodynamic waves (beyond some maximum
wavelength) from which the electrons scatter.
Reynolds (1998) shows that all young SNRs should show some
evidence of nonthermal X-rays due to shock acceleration, but that the maximum
ratio of X-ray/radio nonthermal emission will occur when the SNR age is roughly
equal to the loss time for electrons in the postshock magnetic field -- a
condition that apparently holds for SN~1006 but not for other historical SNRs. 
The nonthermal emission persists to ages beyond $10^4$~yr.

A similar scenario for the nonthermal emission in G347.3$-$0.5 presents some
difficulties. 
We have argued that the distance to this SNR is of
order 6~kpc and that it appears to be in the vicinity of a molecular cloud
complex. The
conditions are thus considerably different than the high-latitude, Type Ia
environment for SN~1006. However, if G347.3$-$0.5 were 
the result of an explosion
in a stellar wind bubble, there are at least qualitative similarities.
The surrounding medium would have low density and a toroidal magnetic field
which may yield conditions for efficient particle acceleration.
Tenario-Tagle et al.~(1991) have performed 2D numerical hydrodynamical
simulations of 
young Type II SNRs interacting with wind-blown bubbles and find that the freely
moving ejecta undergo only partial thermalization while the remnant as a whole
ages rapidly since early evolutionary phases are essentially accelerated due to
the low density medium in the interior of the bubble. If, upon encountering the bubble wall, diffusive shock
acceleration is enhanced, the under-thermalization of the ejecta along with the
small swept-up mass may result in a large ratio of nonthermal/thermal X-ray
emission. Thus, if the remnant is indeed interacting with the observed molecular
clouds, this could enhance the nonthermal component. Simulations indicate
that such interactions result in additional shocks in the cloud which can
enhance the nonthermal electron population (Jun \& Jones 1999) although Jones
\& Kang (1993) argue that such enhancements occur primarily when the incident
shock is already dominated by cosmic-ray pressure.

An alternative interpretation for the featureless X-ray spectrum from
G347.3$-$0.5 could be thermal emission from material dominated by low-$Z$
elements (H, He, C, O) which are fully ionized (Hamilton, Sarazin, 
Szymkowiak, \& Vartanian 1985). Hamilton, Sarazin, \& Szymkowiak (1986)
proposed a scenario for SN~1006 in which the ejecta from the Type Ia explosion
were stratified in such a way as to yield an outer region dominated by such
featureless emission. In addition they were able to obtain an
approximately power-law spectral form by integrating over the power-law
distribution of thermal components that came out of their dynamical model.
However, Koyama et al.~(1995) argue convincingly that their 
ASCA observations of SN~1006 rule this model out. A similar
scenario has been revisited by 
Laming (1998) who concludes that the flux associated with the featureless
components of SN~1006 and G347.3$-$0.5, if interpreted as thermal in
nature, would imply an X-ray emitting mass of low-$Z$ elements far in excess
of that expected to be ejected by a Type Ia supernova explosion. 
Thus, while this mechanism could
potentially explain some small featureless components (\eg for IC443 or Cas A)
it does not appear to be a viable explanation for the spectrum of
G347.3$-$0.5.
Recent model calculations by Baring et al.~(1999) suggest that the
soft X-ray flux of young remnants with parameters typical of Type Ia
explosions should be dominated by bremsstrahlung emission. However,
these authors do find models that produce spectra dominated by
synchrotron emission up to and beyond 10~keV.  One of these
(their model ``D'') uses $E_{51} = 10$,
ambient density $n_0 = 10^{-3} {\rm cm^{-3}}$, ejecta mass $M_{ej}
= 10 M_\odot$, and an age of 40,000 yrs. With the exception of the 
explosion energy (which is probably unrealistically large anyway), these
parameters are moderately consistent with a picture in which G347.3$-$0.5
is a moderate age SNR which is in a low density environment.

The characteristics of the unidentified X-ray source
1WGAJ1713.4$-$3949 are marginally consistent with those for a neutron star. 
The high X-ray to optical flux
ratio virtually rules out a stellar counterpart but is consistent with the
source being a neutron star. The lack of detected pulsations at
current sensitivities does not provide a strong discriminator in the interpretation.  
The spectral characteristics of the 
blackbody model provide a curious scenario if accurate, however. The luminosity
implied by the spectral fit is considerably higher than expected for hot 
polar cap models unless the neutron star is spinning with a period of $\lesssim 
6$~ms. With the rather sparse spectrum, it is conceivable that an additional 
spectral component (e.g., softer blackbody emission from the entire surface) is 
contributing to the somewhat high flux. Alternatively, as mentioned earlier, 
the source could be an extragalactic background object unrelated to
the SNR. More sensitive X-ray measurements are needed to address this.

If 1WGAJ1713.4$-$3949 is indeed a neutron star associated with G347.3$-$0.5, 
then G347.3$-$0.5 differs from SN~1006 in a very fundamental way in 
that the latter is the remnant of a Type Ia explosion.
Association with the molecular cloud seen in CO would lead to a similar 
conclusion; the remnant presumably then originated from a high mass progenitor 
which, in its short lifetime, did not migrate far from its birthplace.  
Models for cosmic ray production 
from shell-type remnants would thus need to accommodate the very different
environments expected from the two types of progenitors. In this case, we 
expect the evolution of G347.3$-$0.5 to have initially been dominated by the 
effects of a precursor wind cavity. If our preferred distance of 6 kpc, as suggested by the apparent 
association with the molecular cloud complex, is correct, then the remnant now has a radius of 
over 50~pc which is still well within the expected size of the stellar wind bubble for massive 
O-type stars (Chevalier \& Liang 1989). This could provide a logical 
explanation for both the low density inferred from the lack of thermal X-ray 
emission as well as the presence of a molecular cloud interaction site.

\section{CONCLUSIONS}

The strong nonthermal X-ray emission observed from G347.3$-$0.5 provides 
tantalizing evidence for the second case of a direct signature of cosmic ray
acceleration in SNRs. Like SN~1006, this remnant is very faint in the radio.
The peak radio surface brightness of SN~1006 is $\sim 90 {\rm\ mJy\ beam}^{-1}$
at the same frequency and resolution, which is comparable to Arc 2 of
G347.3$-$0.5.
However, in distinct contrast, G347.3$-$0.5 shows no evidence of the 
thermal X-ray
emission typically associated with ISM and ejecta material heated by the SNR
shock, nor is the morphology as distinctly shell-like as SN~1006.

We have shown that nonthermal X-ray emission pervades the entire image
of G347.3$-$0.5.  Faint radio emission appears to delineate
a nearly complete shell with brighter arc-like features along regions
of bright X-ray emission. CO observations suggest that these features 
coincide with a region in which the SNR may be interacting with 
a molecular cloud
found along the line of sight. Several lines of evidence, including an
association with this molecular cloud complex as well as an \Hii\ region,
suggest that the most appropriate distance to the remnant is 6 kpc.
The lack of thermal 
emission sets a strong limit on the mean density around the remnant:
$n_0 < 0.28 D_6^{-1/2}\, \rm cm^{-3}$. Although this is rather low for
the near vicinity 
of a molecular cloud, we argue that it means the majority of the remnant is
evolving in the low density interior of the large circumstellar
cavity driven by the wind of its massive 
progenitor star.  Simple Sedov-phase models for the remnant's
evolutionary state can be found that are consistent with this low density
environment for reasonable values of the age and explosion
energy. However in the absence of an actual
measurement of the thermal properties of the
remnant, any conclusions about the evolutionary state of the remnant
remain premature.
In the X-ray bright region toward the northwest edge
of G347.3$-$0.5, where an interaction with
denser gas may be occurring, the limits on the ambient density are
higher $n_o < 1D_6^{-1/2}\, \rm cm^{-3}$ and broadly consistent with the
densities estimated around other SNRs that are associated with
molecular clouds (e.g., N132D in the Large Magellanic Cloud, see
discussion in Hughes, Hayashi, and Koyama 1998).
Regardless of its precise evolutionary state, our
observations reveal G347.3$-$0.5 to be the most extreme example known
of a nonthermal shell-type X-ray remnant and may indicate that shock
acceleration of cosmic rays by supernova remnants can occur in a
diversity of interstellar environments.

Additional measurements will help shed light on the nature of G347.3$-$0.5. In
particular, a search for OH maser emission (Frail et al.~1996) along
the NW limb could confirm a molecular cloud interaction there.
Optical observations of this region could
also provide evidence of such an interaction, but the extinction is rather
high making such observations difficult. Deeper X-ray observations which might
reveal and characterize the thermal component are needed, as are additional
observations of the centrally-located source in order to assess the potential
for association as the compact relic of the explosion which produced
G347.3$-$0.5. While the X-ray spectrum of the SNR clearly indicates the presence
of very high energy electrons, and by inference the acceleration of cosmic
rays, further study is required to provide a sufficiently constrained picture
of G347.3$-$0.5 to which acceleration models can be compared. Higher-energy
X-ray observations to search for any change in the spectral index 
which might be
associated with a bremsstrahlung component will be important in this regard, as
will higher-resolution spatial and spectral observations to provide
additional information on the shock structure, and more sensitive observations
to search for thermal emission.

\acknowledgments

We gratefully acknowledge T.~Handa and T.~Hasegawa of the University
of Tokyo who kindly shared their CO(2--1) observations in the vicinity
of G347.3$-$0.5 prior to publication.  The MOST is operated by the
University of Sydney with support from the Australian Research Council
and the Science Foundation for Physics within the University of
Sydney. BMG acknowledges the support of an Australian Postgraduate
Award. This work was supported in part by the National Aeronautics and
Space Administration through contract NAS8-39073 and grant
NAG5-4803. JPH acknowledges support through NASA grants NAG 5-4794 and
NAG 5-4871.

\end{document}